\documentclass{cimento}
\usepackage{graphicx}
\usepackage{epsfig}

\title{Lowering the Light Speed Isotropy Limit\ETC:
European Synchrotron Radiation Facility Measurements}

\instlist{
\inst{ins:YerPhi}  Yerevan Physics Institute, 375036 Yerevan, Armenia
\inst{ins:IN2P3}   IN2P3, Laboratory for Subatomic Physics and Cosmology, 38026 Grenoble, France
\inst{ins:INFN-1}  INFN sezione di Roma II and Universit\`a ``Tor Vergata", 00133 Roma, Italy
\inst{ins:INFN-2}  INFN sezione di Catania and Universit\`a di Catania, 95100 Catania, Italy
\inst{ins:INFN-3}  INFN sezione di Genova and Universit\`a di Genova, 16146 Genova, Italy
\inst{ins:IN2P3-6} IN2P3, Institut de Physique Nucl\'eaire, 91406 Orsay, France
\inst{ins:INFN-4}  INFN sezione di Torino  and Universit\`a di Torino, 10125 Torino, Italy
\inst{ins:INFN-5}  INFN sezione di Roma I and Istituto Superiore di Sanit\`a, 00161 Roma, Italy
\inst{ins:INSM}    Institute for Nuclear Research, 117312 Moscow, Russia
\inst{ins:INFN-6}  INFN Laboratori Nazionali di Frascati, 00044 Frascati, Italy
\inst{ins:RRC}     RRC, ``Kurchatov Institute", 123182 Moscow, Russia
}

\author{V.G.~Gurzadyan\from{ins:YerPhi}, J.-P.~Bocquet\from{ins:IN2P3},
A.~Kashin\from{ins:YerPhi}, A.~Margarian\from{ins:YerPhi}, O.~Bartalini\from{ins:INFN-1}\ETC,
V.~Bellini\from{ins:INFN-2}, M.~Castoldi\from{ins:INFN-3}, A.~D'Angelo\from{ins:INFN-1},
J.-P.~Didelez\from{ins:IN2P3-6}, R.~Di~Salvo\from{ins:INFN-1}, A.~Fantini\from{ins:INFN-1},
G.~Gervino\from{ins:INFN-4}, F.~Ghio\from{ins:INFN-5}, B.~Girolami\from{ins:INFN-5}, 
A.~Giusa\from{ins:INFN-2}, M.~Guidal\from{ins:IN2P3-6}, E.~Hourany\from{ins:IN2P3-6}, S.~Knyazyan\from{ins:YerPhi}, V.~Kouznetsov\from{ins:INSM}, R.~Kunne\from{ins:IN2P3-6}, 
A.~Lapik\from{ins:INSM}, P.~Levi~Sandri\from{ins:INFN-6}, A.~Lleres\from{ins:IN2P3}, S.~Mehrabyan\from{ins:YerPhi}, D.~Moricciani\from{ins:INFN-1}, V.~Nedorezov\from{ins:INSM}, C.~Perrin\from{ins:IN2P3}, D.~Rebreyend\from{ins:IN2P3}, G.~Russo\from{ins:INFN-2}, 
N.~Rudnev\from{ins:INSM}, C.~Schaerf\from{ins:INFN-1}, M.-L.~Sperduto\from{ins:INFN-2}, M.-C.~Sutera\from{ins:INFN-2}, A.~Turinge\from{ins:RRC}
}

\begin{document}
\maketitle

\begin{abstract}
The measurement of the Compton edge of the scattered electrons in GRAAL facility in 
European Synchrotron Radiation Facility (ESRF) in Grenoble with respect to the Cosmic Microwave 
Background dipole reveals up to 10$\sigma$ variations larger than the statistical errors. 
We now show that the variations are not due to the frequency variations of the accelerator. 
The nature of Compton edge variations remains unclear, thus outlining the 
imperative of dedicated studies.
\end{abstract}

The inverse Compton scattered electron energy spectrum, as suggested in \cite{GM}, provides
a way to study the light speed anisotropy within the energy scales and monochromaticity
reachable in existing synchrotrons. The study of such anisotropy vs the frame, when the dipole 
of the Cosmic Microwave Background (CMB) radiation is vanishing i.e. with respect to the apex 
of the CMB dipole, is of particular interest. It is due to the 'absolute' content of the CMB 
frame within the hierarchy of the motions of the Earth, of the Sun in the Galaxy, of the 
Galaxy in the Local Group, of the Local Group in the Virgo Supercluster \cite{RG}. The 
position of the apex, known since the COBE satellite in 1992, now is available with higher 
precision due to the Wilkinson Microwave Anisotropy Probe (WMAP) 3-year data  \cite{wmap3}. 
Recent indications of the dark energy and stipulated studies of cosmological models with 
varying physical constants, of the dark energy frame and its relation to those of CMB and 
of matter flows, increase the interest to the issue.  

The idea \cite{GM} was to use the fact that, at Compton scattering of a photon of energy $\epsilon$, 
the energy of the scattered photon at small angle scatterings, is
$$
\epsilon ' =  \gamma^2 \epsilon  f (\gamma)  
$$
i.e. scales as the square of the $\gamma$-factor, since $f(\gamma)$ is a weak function of $\gamma$. 
The small angle scatterings are defining the Compton edge (CE), i.e. the upper energy limit for the 
scattered photons directly depending on the electron beam energy.
The high value of $\gamma$ determines the accuracy of $\beta$:
$$
\beta d\beta = (1/\gamma^2) d\gamma/\gamma,
$$
and hence, in the velocity of the light.

The data of highly monochromatic electron beams of the European Synchrotron Radiation Facility (ESRF) 
Compton scattered on laser photons at GRAAL facility \cite{Bo} have been used to study the daily 
variations, i.e. the anisotropy vs the dipole of the Cosmic Microwave Background (CMB) radiation. 
The ESRF parameters are: electron beam of mean energy 6.04 GeV i.e. $\gamma=11820$, scattered on 
laser photons of visible, 514.5 nm, and of several UV lines near 351.1 nm wavelength. The resulting 
CE is at 1100 MeV and 1500 MeV, respectively, for visible and UV photons. The measurements were 
carried on with microstrip detector, so that the accuracy of the CE position given by the microstrip 
is linked to the beam energy.

The analysis of GRAAL data of 1998-2002 (non-continuous) measurements enabled us to obtain a 
conservative upper limit for the anisotropy $$\delta c/c =3\,\, 10^{-12},$$ \cite{mpla} (cf. \cite{Her,Ant}). 
The analysis has been performed following daily variations of the Compton edge vs the azimuthal, declination, as 
well as the angle of the beam vector and the direction of the CMB dipole apex. 

The study of the GRAAL data revealed \cite{mpla}, however, also remarkably high, up to 10$\sigma$
variations in terms of the statistical error bars. In Figs. 1a, 2a we give those data averaged over 
certain time period and angle. No systematic effect has been found responsible for those variations 
and the problem, as stated in \cite{mpla} needed further studies. At the same time, correlations with 
the hours of the beam injection (Fig. 4 in \cite{mpla}), and no correlations with the magnetic dipole 
field variations were noticed.

Continuing the search of systematics, here we represent the results of the analysis of the frequency 
data of the accelerator for a period in June-July 2005 (see the Table 1), as compared with the 
variations of the Compton edge. The Fig. 1a,2a represent the angular distribution in hours and in 
angles, respectively, of all CE data for the periods given in Table 1. CE direction/hour depended 
behaviour (Fig 1a, 2a, 3) remains robust also when various subsets of the data (for various periods) 
are analysed separately, as well as the for the each of UV and optical lines (for details see 
\cite{mpla}). The behaviour of the frequency variations (Fig. 1b, 2b) as seen, shows no correlation 
neither with CE angular (with respect to CMB dipole) nor hour variations.

Thus, we conclude that the frequency variations are not the reason for the CE variations vs CMB 
dipole apex found in \cite{mpla}, further outlining the importance of the use of Compton edge method
at dedicated accelerator measurements. 

%\section*{Acknowledgments}
\stars
ESRF teams, in particular, L.Hardy, J.L.Revol, D.Martin and the members of the alignment group are 
thanked for their help.

\begin{center}
\begin{table}
{
\renewcommand{\baselinestretch}{1.3}
\renewcommand{\tabcolsep}{3.5mm}
\small
\begin{tabular}{lccrr}
\hline
Data          && Time                    & Duration,& Quantity\\
blocks        && of supervision          & day      &         \\
\hline
\hline
CE  I         && 10.04.1998 - 11.05.2002 & 1493	    &    2075 \\
CE II         && 10.04.1998 - 26.11.2002 & 1692	    &    2436 \\
CE III        && 05.06.1999 - 19.04.2001 &  685     &     294 \\
Frequency IV  && 20.06.2005 - 14.07.2005 &	 25     &   26821 \\
CE V          && 01.07.2005 - 11.07.2005 &   11     &     153 \\
\hline
\hline
\end{tabular}
}
\end{table}
\end{center}

\def\figsubcap#1{\par\noindent\centering\footnotesize#1}
\begin{figure}[t]%
\begin{center}
\parbox{2.5in}{\epsfig{figure=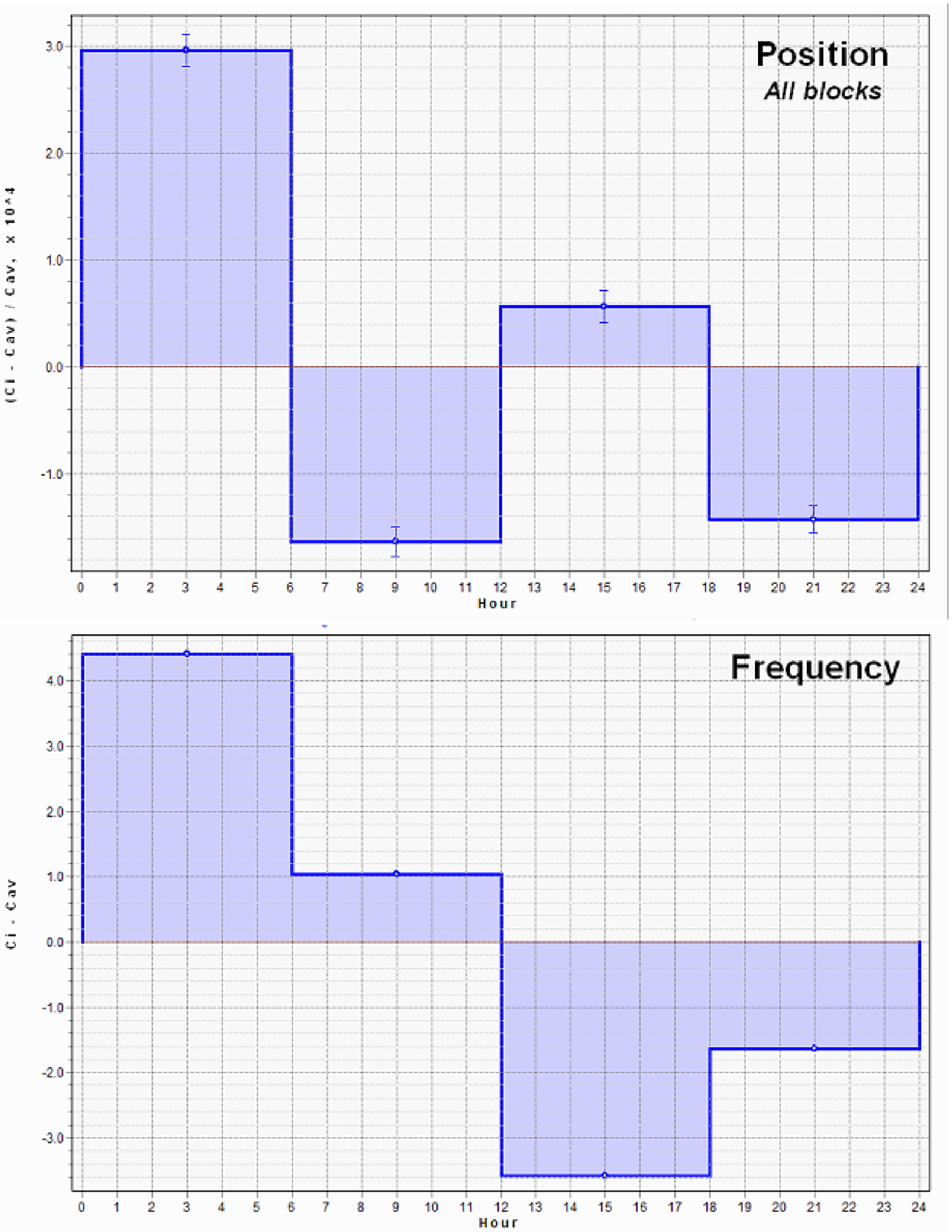,width=2.5in}\figsubcap{Fig. 1. Compton edge (a., upper plot) and accelerator frequency (b., lower) daily variations, data averaged in 6-hour intervals. Note, the small error bars for CE. }}
\hspace*{5pt}
\parbox{2.5in}{\epsfig{figure=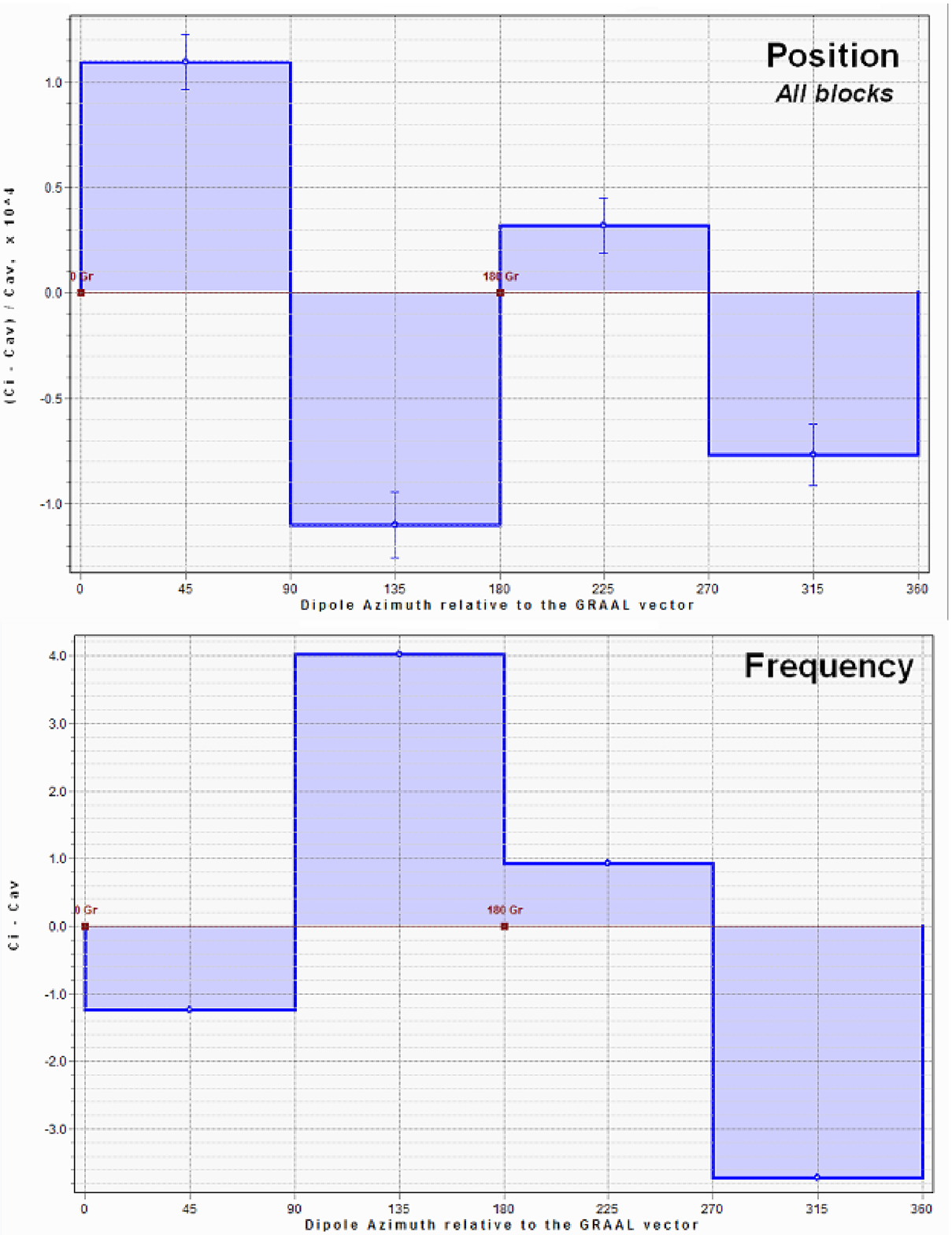,width=2.5in}\figsubcap{Fig. 2. The same as in Fig 1, now 
vs the azimuth of the CMB dipole. The inconsistency of CE and accelerator frequency variations is visible.}}
\label{aba:fig1}
\end{center}
\end{figure}

\begin{figure}[b]%
\begin{center}
\psfig{file=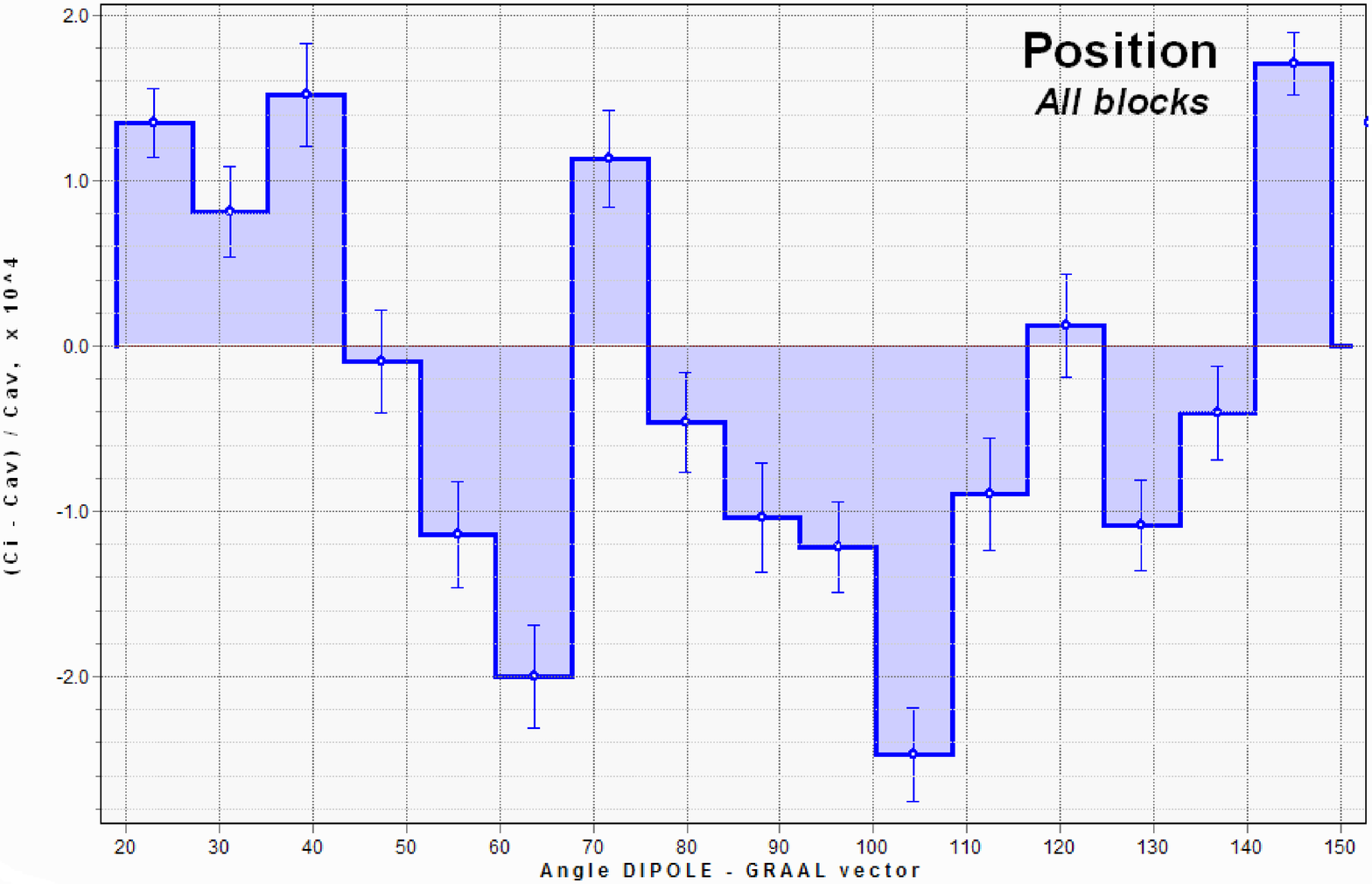,width=4in}\figsubcap{Fig. 3. Compton edge anisotropy variations 
vs the angle between the CMB dipole and GRAAL beam vector; averaging over $8.1^{\circ}$.}
\end{center}
\label{aba:fig1}
\end{figure}

\end{document}